\documentclass[
  journal=medium,
  manuscript=article-type,
  year=2020,
  volume=37,
]{cup-journal}

\usepackage{amsmath}
\usepackage[nopatch]{microtype}
\usepackage{booktabs}

\title{Advanced Techniques in Mortality Trend Estimation: Integrating Generalized Additive Models and Machine Learning to Evaluate the COVID-19 Impact}

\author{Asmik Nalmpatian}
\affiliation{Department of Statistics, LMU Munich, Germany}
\email[Asmik Nalmpatian]{asmik.nalmpatian@campus.lmu.de}

\author{Christian Heumann}
\affiliation{Department of Statistics, LMU Munich, Germany}

\author{Stefan Pilz}


\pdfoutput=1
  
\begin{document}

\begin{abstract}
The last two centuries have seen a significant increase in life expectancy. Although past trends suggest that mortality will continue to decline in the future, uncertainty and instability about the development is greatly increased due to the ongoing COVID-19 pandemic. It is therefore of essential interest, particularly to annuity and life insurers, to predict the mortality of their members or policyholders with reliable accuracy.  
The goal of this study is to improve the state-of-the-art stochastic mortality models using machine learning techniques and generalize them to a multi-population model. Detailed cross-country results conducted for Finland, Germany, Italy, the Netherlands, and the United States show that the best forecasting performance is achieved by a generalized additive model that uses the framework of APC analysis. Based on this finding, trend forecasts of mortality rates as a measure of longevity are fulfilled for the future, given a range of COVID-19 scenarios, from mild to severe.  
Discussing and evaluating the plausibility of these scenarios, this study is useful for preparation, planning, and informed decision-making.
\end{abstract}

\noindent  

\section{Introduction}
In recent decades, life expectancy in the developed world has shown a significant increase, exemplified by a notable 56.7\% reduction in the mortality rate of 80-year-old men in the USA from 1933 to 2019. This decline has spurred interest from governments, private pension companies, and life insurers to develop models and projections for future mortality trends. Accurately capturing and forecasting these trends is essential for informed decision-making, particularly amidst the uncertainties heightened by the COVID-19 pandemic. Proper prediction of population ageing is crucial to ensure the sustainability of pension schemes, life insurers, and social security systems. 
Since the onset of the COVID-19 pandemic, numerous contributions have been made to better understand and predict the diverse implications of this disease. In this study, we specifically focus on assessing the impact of COVID-19 on short to mid-term mortality trends. Several approaches have been explored on our way to identify the most suitable method for addressing this issue, leading us to our first research question: \textbf{How can the performance of traditional mortality models be enhanced in terms of fit and forecast, while still maintaining explainability?}

This research paper introduces two notable improvements to traditional mortality models: Cross-country Generalized Additive Model (GAM) with Age-Period-Cohort (APC) framework in comparison to the classical APC \autocite{clayton1987models} model and a Tree-based Machine Learning (ML) Model \autocite{levantesi2019application} in relation to Lee Carter (LC) model as baseline \autocite{lee1992modeling}. After discussing the literature, the GAM with APC framework is chosen for trend forecasting, while the ML method proves as a reliable diagnostic tool to identify weaknesses in traditional baseline mortality modeling.

The literature highlights several contemporary approaches that aim to enhance mortality models, including the utilization of ML techniques such as the two-step-approach of \cite{levantesi2019application}, as well as the application of neural networks to improve models like the Common-Age-Effect Model introduced by \cite{schnurch2022impact} and the extension of LC for multiple populations as demonstrated by \cite{richman2021neural}.

GAMs are a well-established model class, first introduced by \cite{hastie1987generalized} as an extension of classical Generalized Linear Models \autocite{nelder1972generalized} to incorporate non-parametric smooth components and provide a robust and flexible framework for modeling nonlinear effect structures \autocite{wood2017generalized}. Moreover, GAMs have been applied in various COVID-related modeling tasks. For instance, \cite{fritz2022statistical} utilized GAMs to model the incidence of hospitalizations, taking into account the temporal delay in data availability. Additionally, numerous applications have explored the potentially nonlinear and delayed effects of meteorological factors (such as temperature, humidity, and rainfall) on COVID-19 cases and deaths (\cite{corlett2020impacts}; \cite{prata2020temperature}; \cite{ward2020role}; \cite{zhu2020association}) or COVID-effect on one-country population mortality \autocite{izadi2021generalized}. To summarize, an advantage of GAMs is their potential extension to incorporate additional factors, which other methods like the recalibration of the Li and Lee Mortality Model for the Multi-populational setting lack \autocite{robben2022assessing}.

While APC analyses have traditionally originated in epidemiological science \autocite{kupper1985statistical}, there has been a growing utilization of APC methods across various research fields in recent decades \autocite{yang2013age}. Spline-based regression has emerged as a popular model class in APC analysis since the late 1990s, allowing for the estimation of potentially nonlinear age, period, and cohort effects (e.g. \cite{carstensen2007age}; \cite{heuer1997modeling}). Building upon this approach, \cite{clements2005lung} proposed an APC model that utilizes a bivariate spline function dependent on age and period within a GAM. The resulting two-dimensional interaction surface implicitly captures cohort information along the diagonals and thus addresses the identification problem of the linear dependency of age, period, and cohort (cohort = year - age). Penalized splines, such as those proposed by \cite{eilers1996flexible}, allow for the estimation of nonlinear relationships while avoiding overfitting through the imposition of a penalty on the roughness of the effects.

Our paper extends the current literature in several ways. We apply GAM in the APC framework, using a smoothed second-order spline with penalty points. This allows for cross-country forecasting of future mortality trends and the possible inclusion of further, i.e. socioeconomic variables, which to our knowledge is the first time they have been combined in this manner. This approach will help us examine the ultimate research question: \textbf{How will mortality develop in the future in different countries, considering the COVID-19 impact?} We conduct a detailed examination of Germany, Finland, the Netherlands, Italy (representing Europe), and the United States (representing North America) to compare the above-outlined approaches and conduct forecasts based on four future Scenarios, utilizing data from the Human Mortality Database \autocite{hmd}, supplemented by COVID-related data from Short-term Mortality Fluctuations collected on a weekly basis and aggregated to a yearly level \autocite{stmf}. Age buckets are transformed to a metric individual-aged scale as suggested by \cite{antonio2020assessing}, ensuring consistency with traditional models and distinguishing our approach from other COVID impact modeling efforts (e.g. \cite{schnurch2022impact}). The cross-country approach attempts to model and combine evidence on all data, while still allowing to model country-specific differences in mortality. The paper follows a structured approach, starting with an explanation of the database and the two methods used. The results section compares the performance of both methods across countries, leading to insights into the preferred method for trend forecasting. The paper concludes with a discussion and summary of the findings.

\section{Data and Methods}
\subsection{Data}

The study utilizes the Human Mortality Database \autocite{hmd}, which contains comprehensive data on mortality rates, number of deaths, and population size for various subpopulations based on country and gender. The dataset covers a wide age range, from 0 to 100+ years, and the observed mortality rates are calculated by dividing the number of deaths by the corresponding population size.
Countries were selected based on their geographic context and the contrasting impact of the COVID-19 pandemic, with Finland, Germany, Italy, the Netherlands representing Europe, and the United States representing North America. Data from the Short-Term Mortality Fluctuations \autocite{stmf} series, collected weekly and aggregated into age buckets, were transformed into annual mortality data at individual ages using a methodology described by \cite{antonio2020assessing}.
The analysis extends up to 2021, including years impacted by COVID-19. Traditional mortality and breakpoint detection models may not accurately capture the effects of events like COVID-19 for several reasons. First, these models typically focus on detecting breakpoints within the time range and may not adequately account for breakpoints at the edge of the time range, such as those caused by unprecedented events. As a result, these events could for example be treated as outliers or noise and not considered when forecasting future years. This limitation calls for alternative approaches that can effectively capture and incorporate the impact of such events.

\begin{figure}[hbt!]
\centering
\includegraphics[width=0.9\linewidth]{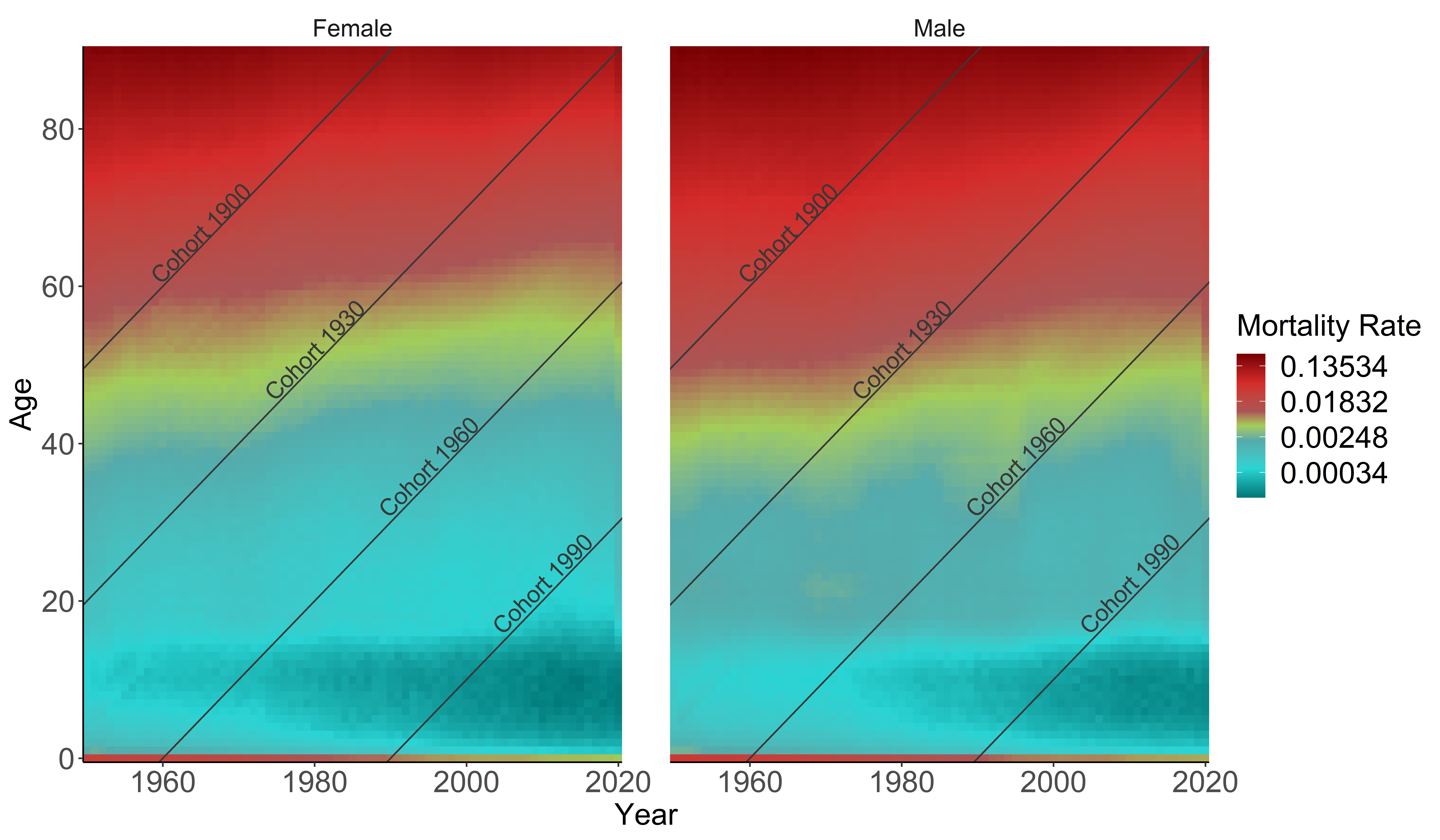}
\caption[Heatmaps of mortality rates for the US population]{Heatmaps of mortality rates for the US population are shown, with age groups and periods represented horizontally and vertically, respectively. The diagonal lines display unique cohorts.}
\label{fig_1}
\end{figure}

Figure \ref{fig_1}'s heatmaps visualize mortality rate changes in the United States, showing a decreasing trend over the years. Darker red colors indicate higher mortality rates, with females generally exhibiting lower rates than males, especially in older age categories. Infant mortality rates have significantly improved, transitioning from red to green. The diagonal lines represent different birth cohorts, providing insights into age, period, and cohort effects on mortality rates, valuable for modeling purposes.

\subsection{Methods} 

The aim of the developed and applied methodology is to enhance the traditional modeling of mortality by focusing on performance in fitting, and particularly in forecasting, while ensuring explainability. The ultimate objective is to select the most suitable and advanced model that can be utilized for trend forecasting to assess the future impact of COVID. In this pursuit, various methods have been tested, with particular emphasis on comparing Tree-Based Machine Learning (ML) and Generalized Additive Models (GAM) approaches to traditional methods.

The link between the ML and the GAM approach lies in their complementary strengths and the advantages they offer in enhancing traditional mortality models. GAMs, originally introduced as an extension of Generalized Linear Models, provide a flexible framework to model nonlinear but smooth effect structures, making them well-suited for capturing complex relationships in mortality data. Moreover, they have demonstrated effectiveness in various COVID-related modeling tasks, such as modeling hospitalizations and exploring the effects of meteorological factors on COVID-19 cases and deaths.

On the other hand, machine learning techniques, specifically the Tree-Based ML methods, offer powerful diagnostic tools to identify weaknesses in traditional baseline mortality modeling by estimating the improvement factors by given variables. The ML approach can provide valuable insights into the data and help enhance the performance of mortality models by uncovering patterns and relationships that might be challenging to detect using traditional statistical methods.

To briefly touch on the background of Random Forests, it is a machine learning algorithm that combines the predictions of multiple decision trees to make more accurate predictions. Decision trees are non-parametric models used for regression and classification tasks, aiming to predict outcomes based on simple decision rules (see more in the appendix). For a given outcome variable $a$ at time $t$ and subpopulation $s$, the Random Forest estimator $q(a,t,s)$ is calculated as the average of the predicted outcomes from the $B$ decision trees: 

\begin{equation}
\begin{aligned}\label{eq:first}
\hat{q}_{a,t,s} = \frac{1}{B}  \sum_{b = 1}^{B} \hat{q}_{a,t,s}^{(b)} 
\end{aligned}
\end{equation}

The Random Forest algorithm employs a two-step process: bagging, which involves creating multiple decision trees from bootstrap samples of the data, and random feature selection, where only a subset of features is considered at each split in the tree. The final prediction is obtained by aggregating the results of all the trees. Gradient Boosting is mentioned as an alternative tree-based machine learning method, and further details regarding its implementation can be found in the appendix.

In this study, a 2-step approach proposed by \cite{levantesi2019application} is followed to improve the Lee Carter (LC) mortality model. The first step involves estimating the mortality rates $\mu_{a,t,s}$ using the LC model (see appendix for more details), while the second step corrects these estimations using a multiplicative factor $q$ to account for underestimations $q>1$ or overestimations $q<1$. The formula used to calculate the corrected mortality rates is explained as follows:

In the first stage, the LC model is fitted to the data for a unique country and gender combination. The estimated mortality rates $\mu_{a,t,s}^{LC}$ are evaluated using this model. In the second stage, three supervised tree-based machine learning models (Decision Tree, Random Forest, and Gradient Boosting Machine) are used to calibrate the improvement factor $q_{a,t,s}$ using age, year, and cohort as input features. The relationship between the observed deaths $D_{a,t,s}$ and the estimated deaths $d_{a,t,s}$ is represented by a Poisson distribution with a multiplicative factor:

\begin{equation}
\begin{aligned}\label{eq:first}
D_{a,t,s} \sim Poisson(d_{a,t,s} \cdot q_{a,t,s}) 
\end{aligned}
\end{equation}
where $d_{a,t,s}$ is the product of the expected population size $E_{a,t,s}$ and the estimated mortality rates $\mu_{a,t,s}^{LC}$.

In the second stage, tree-based machine learning algorithms are applied to estimate $q_{a,t,s}$ as target, considering age, year, and cohort as predictors and $d_{a,t,s}$ as exposure. The improved mortality rates result in $\mu_{a,t,s}^{ML} = \mu_{a,t,s}^{LC} \cdot q_{a,t,s}^{ML}$. To forecast these improved mortality rates, an ARIMA model with automated parameter estimation is selected \autocite{hyndman2008automatic}. \\

The APC framework is utilized in the next method, considering age, year, and cohort as predictors for modeling mortality rates with GAM – in a multipopulational cross-country approach. The linear combination of these predictors may pose identification challenges, which are explicitly addressed in classical APC models. The GAM framework with a bivariate spline function is proposed by \cite{clements2005lung}, incorporating age and period to create a two-dimensional interaction surface that implicitly captures cohort information on its diagonals. Penalized B-splines are used to define the marginal spline bases, ensuring accurate modeling of non-linear relationships while avoiding overfitting. The semiparametric additive Poisson regression with a log link function and offset for population size $E_{a,t,s}$ is applied to model aggregated death counts and mortality rates: with a model structure formulated as follows: 

\begin{equation}
\begin{aligned}\label{eq:first}
\log(\mu_{apc,i}) = \beta_0 + f_{ap,s}(age_i, period_i) + \log(E_{i})
\end{aligned}
\end{equation}

Here, $f_{ap,s}$ represents a bivariate function of age $a$ and period $p$ specific to each country and gender combination $s$; $i$ represents any observation unique to a particular combination of age, time period, gender, and country. It is modeled using a tensor product of spline bases for age and period. The cross-country model estimates a separate bivariate function for each country and gender, capturing the specific mortality patterns within each subpopulation and at the same time allowing countries to learn from each other’s experience in the intercept term.

The forecast of mortality rates is based on extrapolation of the spline fit, assuming a globally quadratic structure and a persistent curvature outside the observed data. The choice of degrees of freedom for the covariates can be either predetermined or estimated automatically. We caution against extrapolating too far into the future.

To visualize the marginal effects of each component in an accessible way, temporal developments are condensed in one specific dimension (either age or period) and averaged over the respective other component. This allows for examining the effects of age or period while considering the cohort values as post-stratification (\cite{weigert2022semiparametric}; \cite{wood2017generalized}):

\begin{equation}
\begin{aligned}\label{eq:first}
f_a(age) &= \frac{1}{{P}} \sum_{period \in P} f_{ap}(period{\vert}age)  \\ 
f_p(period) &= \frac{1}{{A}} \sum_{age \in A} f_{ap}(age{\vert}period)   \\ 
f_c(cohort) &= \frac{1}{{A} \cdot {P}} \sum_{age \in A} \sum_{period \in P} {ap}(age, period{\vert}cohort) 
\end{aligned}
\end{equation}

\section{Results}
\subsection{Comparison of Tree-Based ML techniques and traditional mortality models}

In our two-stage approach, we first employed the classical Lee Carter (LC) model and then enhanced it using tree-based machine learning with a multiplicative improvement factor. The machine learning algorithms (Decision Tree, Random Forest, and Gradient Boosting) were implemented using the \texttt{rpart}, \texttt{randomForest}, and \texttt{gbm} packages, respectively (\cite{therneau2015package}; \cite{rcolorbrewer2018package}; \cite{ridgeway2004gbm}), with specific parameter settings based on recommendations provided by \cite{levantesi2019application}. 

\begin{figure}[hbt!]
\centering
\includegraphics[width=0.75\linewidth]{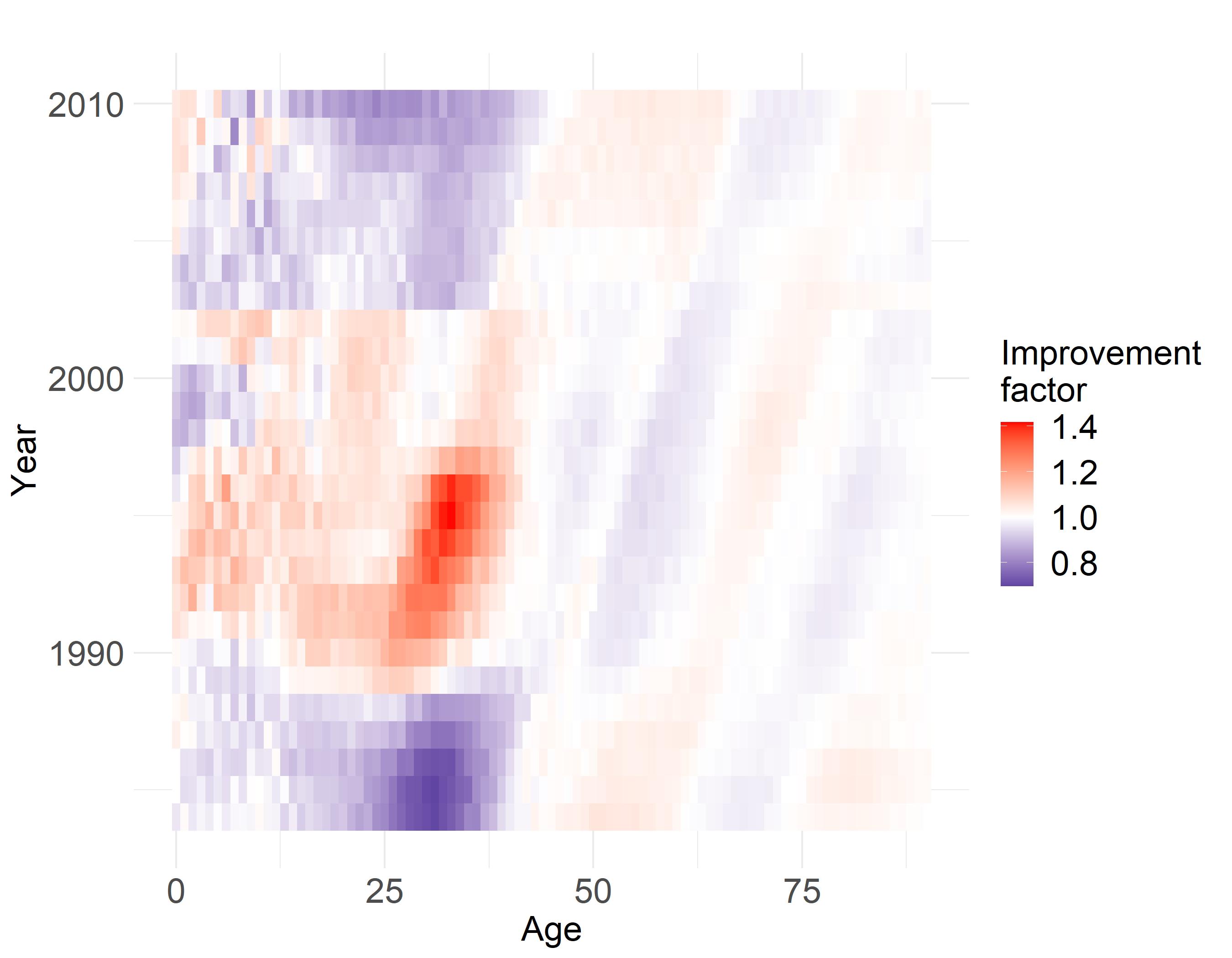}
\caption[Tree-based improvement factors for the Italian males]{Heatmaps of tree-based improvement factors are presented, showing values for each year (1950-2010) and age (0-90) using RF model for the Italian male population.}
\label{fig_2}
\end{figure}

 The color-coded heatmap in figure \ref{fig_2} illustrates the magnitude of the improvement factors $q_{a,t,s}$, with white areas indicating a value of 1, suggesting a perfect fit of the mortality rates according to the LC model. Areas with values smaller than 1 (purple) suggest that the mortality model may overestimate the rates for example for middle ages before 1990, while values higher than 1 (dark orange) suggest underestimation for middle aged between 1990 and 2000. The diagonal splits in many areas highlight the cohort effect, justifying the inclusion of the cohort parameter in the tree-based algorithms. This visualization demonstrates the local variations and improvements achieved by incorporating tree-based machine learning techniques into the mortality modeling process.

To evaluate the accuracy of the models, the root-mean-square error (RMSE) was calculated, which measures the average difference between the predicted number of deaths $d_{a,t,s}$ and the actual observed number of deaths $D_{a,t,s}$ across all age $a$, year $t$, and subpopulation $s$ combinations. The RMSE is computed as follows:

\begin{equation}
\begin{aligned}\label{eq:first}
RMSE_{a,t,s} = \sqrt{\frac{\sum_{s \in S} \sum_{a \in A} \sum_{t \in T}  (d_{a,t,s} - D_{a,t,s})^2 }{n}}
\end{aligned}
\end{equation}

where $n$ represents the total number of data points.

\begin{table}[hbt!]
\begin{threeparttable}
\caption[In-sample RMSE for LC and the tree-based improvements]{Table displays the in-sample RMSE for LC and tree-based improvements using Decision Tree, Random Forest, and Gradient Boosting. The analysis covers multiple countries, genders, and time periods, with the fitted period spanning from 1950 to 2010 for Finland, Italy, the Netherlands, and the US, and from 1990 to 2010 for Germany.}
\label{table_1}
   \centering
    \footnotesize 
    \setlength\tabcolsep{1.5pt}
    \begin{tabular}{c   cccc   cccc} 
        \toprule
        \multicolumn{1}{c}{Country} & \multicolumn{4}{c}{Female} & \multicolumn{4}{c}{Male} \\
        
        \cmidrule(lr){2-5}\cmidrule(lr){6-9} 
        
        & \multicolumn{1}{c}{LC}  & \multicolumn{1}{c}{Tree}  & \multicolumn{1}{c}{RF}  & \multicolumn{1}{c}{GBM} & \multicolumn{1}{c}{LC}  & \multicolumn{1}{c}{Tree}  & \multicolumn{1}{c}{RF}  & \multicolumn{1}{c}{GBM} \\

        \cmidrule(lr){1-1}\cmidrule(lr){2-2}\cmidrule(lr){3-3}\cmidrule(lr){4-4}%
        \cmidrule(lr){5-5}\cmidrule(lr){6-6}\cmidrule(lr){7-7}\cmidrule(lr){8-8}%
        \cmidrule(lr){9-9}%
        FIN  &  0.0045 & 0.0021 & 0.0027 & 0.0011 & 0.0072   & 0.0019 & 0.0033 & 0.001 \\
        DE   &  0.0015 & 0.0014 & 0.0006 & 0.0004 & 0.0021 & 0.0019 & 0.0008 & 0.0006 \\
        ITA  &  0.0025 & 0.0007 & 0.0019 & 0.0008 & 0.0012 & 0.0006 & 0.0008 & 0.0003 \\
        NLD  &  0.0019 & 0.0004 & 0.0011 & 0.0002 & 0.0017 & 0.0000 & 0.0008 & 0.0000 \\
        US   &  0.0014 & 0.0007 & 0.0012 & 0.0005 & 0.0017 & 0.0012 & 0.0014 & 0.0006 \\
        \bottomrule
    \end{tabular}  
\end{threeparttable}
\end{table}

Table \ref{table_1} provides a detailed analysis of the goodness-of-fit, showing significant reductions in RMSE for all machine learning methods compared to the original LC model, with GBM achieving the highest reduction for in-sample predictions. The graphs also illustrate how machine learning estimators improved the fit of the mortality model. Additionally, out-of-sample forecasting was conducted using the improved models, as shown in Table \ref{table_2}, demonstrating the ability to generalize and assess the forecasting quality. While the reduction in RMSE was not as large as in the in-sample expansion case, the trees showed better resolution of locality. However, overfitting issues were observed, particularly in Decision Trees, highlighting the bias-variance dilemma. Random Forests proved effective in overcoming overfitting and achieved the largest decrease in RMSE. Overall, applying machine learning techniques significantly improved the fit and forecast accuracy compared to the original LC model, providing also possibilities to understand the derived solutions, i.e. visually.

\begin{table}[hbt!]
\begin{threeparttable}
\caption[Out-of-sample RMSE for LC and the tree-based improvements]{Table displays the out-of-sample RMSE for LC and tree-based improvements using Decision Tree, Random Forest, and Gradient Boosting. The analysis covers multiple countries, genders, and time periods, with the fitted period spanning from 1950 to 2010 for Finland, Italy, the Netherlands, and the US, and from 1990 to 2010 for Germany. The out-of-sample years are 2011-2019.}
\label{table_2}
   \centering
    \footnotesize 
    \setlength\tabcolsep{1.5pt}
    \begin{tabular}{c   cccc   cccc} 
        \toprule
        \multicolumn{1}{c}{Country} & \multicolumn{4}{c}{Female} & \multicolumn{4}{c}{Male} \\
        
        \cmidrule(lr){2-5}\cmidrule(lr){6-9} 
        
        & \multicolumn{1}{c}{LC}  & \multicolumn{1}{c}{Tree}  & \multicolumn{1}{c}{RF}  & \multicolumn{1}{c}{GBM} & \multicolumn{1}{c}{LC}  & \multicolumn{1}{c}{Tree}  & \multicolumn{1}{c}{RF}  & \multicolumn{1}{c}{GBM} \\

        \cmidrule(lr){1-1}\cmidrule(lr){2-2}\cmidrule(lr){3-3}\cmidrule(lr){4-4}%
        \cmidrule(lr){5-5}\cmidrule(lr){6-6}\cmidrule(lr){7-7}\cmidrule(lr){8-8}%
        \cmidrule(lr){9-9}%
        FIN  &  0.0021 & 0.0021 & 0.0014 & 0.0017 & 0.0029 & 0.0034 & 0.0024 & 0.0031 \\
        DE   &  0.0048 & 0.0046 & 0.0032 & 0.0033 & 0.0052 & 0.0052 & 0.0033 & 0.0034 \\
        ITA  &  0.0045 & 0.0046 & 0.0039 & 0.0022 & 0.0042 & 0.0041 & 0.0020 & 0.0022 \\
        NLD  &  0.003  & 0.003  & 0.0015 & 0.0017 & 0.0035 & 0.0035 & 0.0018 & 0.0020 \\
        US   &  0.0023 & 0.0023 & 0.0010 & 0.0011 & 0.0054 & 0.0052 & 0.0049 & 0.0051 \\
        \bottomrule
    \end{tabular}  
\end{threeparttable}
\end{table}

\subsection{Performance evaluation of GAM-based APC improvements}
The predictive power of the GAMs with the APC framework was evaluated using a test dataset that was not included in the model training. The accuracy of the forecasts was assessed by comparing the predicted number of deaths $d_{a,t,s}$ to the actual observed number of deaths $D_{a,t,s}$ for each age $a$, year $t$, and subpopulation $s$. The dataset has been pooled together for all countries, the fitted years for testing purposes ranged from 1990 to 2015, while the testing period encompassed the years 2016 to 2019. The forecasts of the 2016-2019 data were compared to the actual observed deaths, and the RMSE was calculated as a measure of prediction accuracy. The models presented in this study were implemented using the R package \texttt{mgcv} \autocite{wood2015package}.

\begin{table}[hbt!]
\begin{threeparttable}
\caption[In-sample RMSE for APC and the GAM-based improvements]{Table presents the out-of-sample RMSE for different models, including APC and GAM-based improvements using Age-Period tensor product splines, as well as continentwise modeling. The analysis covers multiple countries, genders, and time periods, with the fitted period spanning from 1990 to 2015 for Finland, Germany, Italy, the Netherlands, and the US.}
\label{table_3}
   \centering
    \footnotesize 
    \setlength\tabcolsep{1.5pt}
      \begin{tabular}{c   cccc   cc} 
        \toprule
        \multicolumn{1}{c}{Country} & \multicolumn{3}{c}{Female} & \multicolumn{3}{c}{Male} \\
        
        \cmidrule(lr){2-4}\cmidrule(lr){5-7} 
        
        & \multicolumn{1}{c}{Class. APC}   & \multicolumn{1}{c}{GAM}  & \multicolumn{1}{c}{Continentwise} 
        & \multicolumn{1}{c}{Class. APC}   & \multicolumn{1}{c}{GAM}  & \multicolumn{1}{c}{Continentwise} \\

        \cmidrule(lr){1-1}\cmidrule(lr){2-2}\cmidrule(lr){3-3}\cmidrule(lr){4-4}%
        \cmidrule(lr){5-5}\cmidrule(lr){6-6}\cmidrule(lr){7-7}%
        
       FIN  &  0.0015   & 0.0011 & 0.0013      & 0.0027 & 0.0013   & 0.0015 \\
        
        DE   &  0.0022   & 0.001 & 0.0014        & 0.0021 & 0.001   & 0.0013 \\
        
        ITA  &  0.0021   & 0.001 & 0.0011       & 0.0025  & 0.001   & 0.0011 \\
        
        NLD  &  0.0032   & 0.001   & 0.0011      & 0.0015 & 0.001  & 0.0012 \\
        
        US   &  0.0033   & 0.0013  & 0.0016      & 0.0028  & 0.0011 & 0.001 \\
        
        \bottomrule
    \end{tabular}
\end{threeparttable}
\end{table}

Table \ref{table_3} and \ref{table_4} present the in-sample and out-of-sample results of different models for each country and gender, respectively. The first column represents the classical APC model, while the second column represents the APC model built with GAM using tensor products of age and period. The third column corresponds to the continentwise approach, where separate GAM models were estimated for the EU countries and the USA.

\begin{table}[hbt!]
\begin{threeparttable}
\caption[Out-of-sample RMSE for APC and the GAM-based improvements]{Table presents the out-of-sample RMSE for different models, including APC and GAM-based improvements using Age-Period tensor product splines, as well as continentwise modeling. The analysis covers multiple countries, genders, and time periods, with the fitted period spanning from 1990 to 2015 for Finland, Germany, Italy, the Netherlands, and the US. The out-of-sample years evaluated are from 2016 to 2019.}
\label{table_4}
   \centering
    \footnotesize 
    \setlength\tabcolsep{1.5pt}
      \begin{tabular}{c   cccc   cc} 
        \toprule
        \multicolumn{1}{c}{Country} & \multicolumn{3}{c}{Female} & \multicolumn{3}{c}{Male} \\
        
        \cmidrule(lr){2-4}\cmidrule(lr){5-7} 
        
        & \multicolumn{1}{c}{Class. APC}   & \multicolumn{1}{c}{GAM}  & \multicolumn{1}{c}{Continentwise} 
        & \multicolumn{1}{c}{Class. APC}   & \multicolumn{1}{c}{GAM}  & \multicolumn{1}{c}{Continentwise} \\

        \cmidrule(lr){1-1}\cmidrule(lr){2-2}\cmidrule(lr){3-3}\cmidrule(lr){4-4}%
        \cmidrule(lr){5-5}\cmidrule(lr){6-6}\cmidrule(lr){7-7}%
        
        FIN  &  0.0029 & 0.0012 &  0.0014 & 0.0029 & 0.0015  &  0.0015 \\
        
        DE   &  0.0046 & 0.0021 &  0.0019 & 0.0045 & 0.002   & 0.0023 \\
        
        ITA  &  0.0025 & 0.0016 &  0.0014 & 0.0021 & 0.0013  & 0.0015 \\
        
        NLD  &  0.0020 & 0.0013 &  0.0011 & 0.0038 & 0.0011  & 0.0012 \\
        
        US   &  0.0018 & 0.0010 &  0.0019 & 0.0020 & 0.0016  & 0.0018 \\
        
        \bottomrule
    \end{tabular}
\end{threeparttable}
\end{table}

The out-of-sample RMSE values indicate the forecast errors for each model. It is observed that the GAM-based APC model achieved a significant reduction not only in fit but also in forecast errors compared to the classical APC model. This implies that the GAM approach improves the accuracy of mortality rate predictions.

\begin{figure}[hbt!]
\centering
\includegraphics[width=0.75\linewidth]{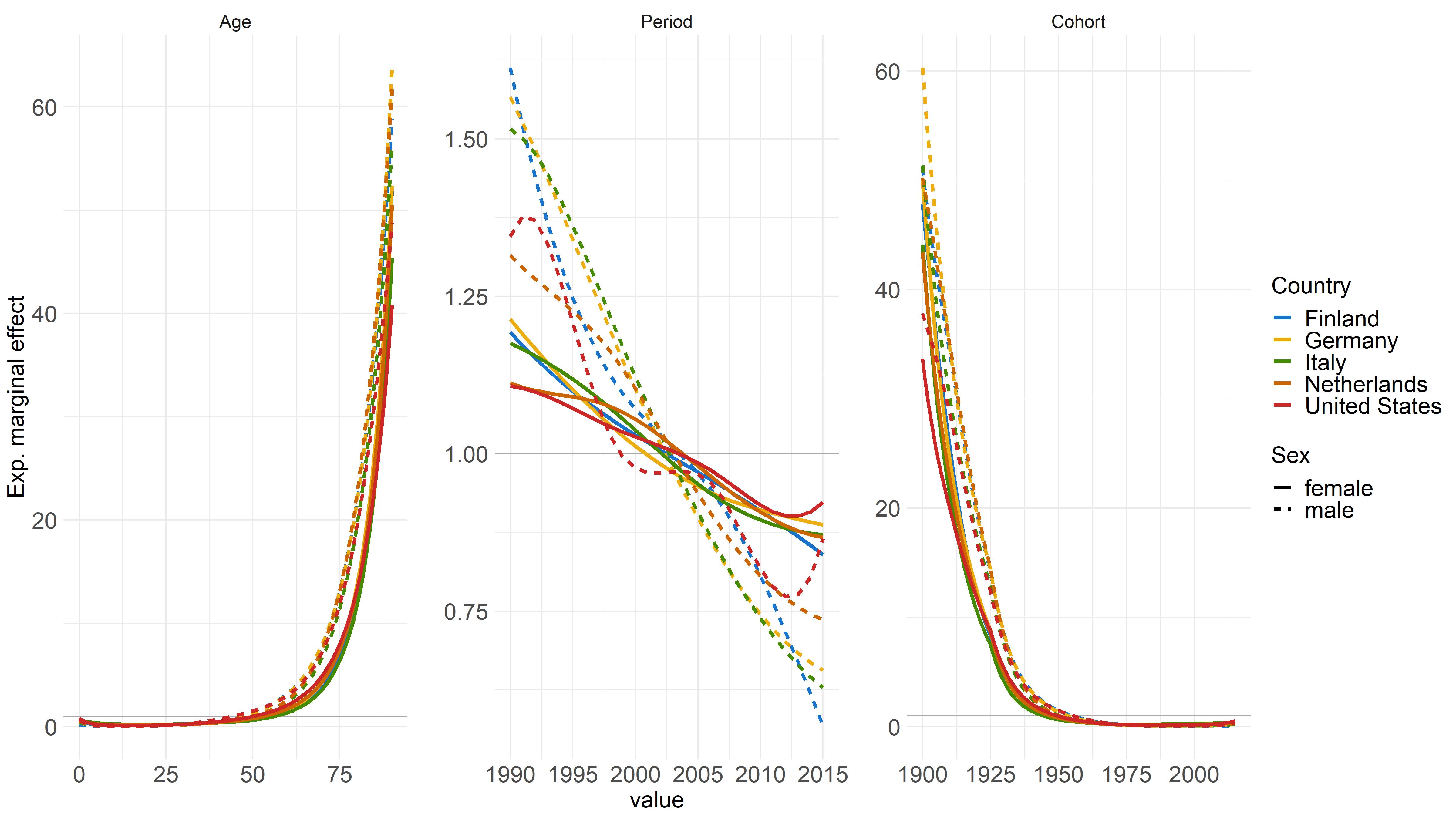}
\caption[Estimated marginal effects per country and gender (1990-2015)]{Estimated marginal effects of age, period, and cohort on mortality rates are depicted in the figure, considering multiple countries and genders. The horizontal lines represent the level of no effect. The GAM model was fitted for the years 1990-2015 and ages 0-90.}
\label{fig_3}
\end{figure}

This model allows for the interpretation of the marginal effects of each variable across different countries and genders as shown in figure \ref{fig_3}. One key highlight is the multipopulational modeling approach, which employs GAMs to capture non-linear effects and enhance the robustness and flexibility of mortality modeling. The model enables the interpretation of exponential marginal effects, with age, period, and cohort being the components analyzed.

Regarding age effects, they represent variations linked to biological and social ageing processes that are specific to individuals (e.g. \cite{reither2009birth}; \cite{carnes1996continuing}; \cite{olshansky1997ever}).

As expected, higher ages are associated with higher mortality rates, and this effect is stronger for males but similar across countries. Cohort effects, on the other hand, reflect variations resulting from the unique experiences of individuals based on their year of birth. (e.g. \cite{hamilton2019age}; \cite{crimmins2019differences}; \cite{trovato1996narrowing}). Similar reverse effects can be observed. 

Period effects, which result from external factors affecting all age groups equally at a given point in time, show interesting patterns (e.g. \cite{rosella2016narrowing}; \cite{perls1998women}).
While the descending trend for women is relatively consistent and shallow across all countries, men exhibit a much steeper decline, indicating a stronger improvement in mortality rates over the years. Notably, there are noticeable increases in mortality rates for Italy and the US in recent years, particularly for US males, which may be associated with factors such as the opioid crisis. 

One advantage of using GAMs is the ability to analyze different countries simultaneously, potentially providing insights into socio-economic factors at work. The choice of GAMs for further analysis is justified based on their superior performance compared to other methods. Both GAMs and tree-based machine learning approaches show improved fit performance, but GAMs demonstrate a stronger improvement in forecast accuracy. Although GAMs have limitations in terms of longer-term forecasts, they are suitable for short-term forecasts within a few years, which aligns with the goal of assessing the impact of COVID-19 in the upcoming 4–5 years.

It is important to note that parallel methods can serve different purposes effectively, and combining tree-based machine learning techniques as a diagnostic tool to detect weaknesses in the base model and incorporate them into GAMs can be beneficial. Overall, the GAM model is leveraged to achieve the ultimate goal of short-term COVID-19 forecasts while considering the scope and focus of the study.

\subsection{Trend forecast considering COVID-19}

Even though the impact of COVID-19 is, fortunately, diminishing in the present time, ints impact on historical (and future) data and the persisting uncertainties in the future cannot be overlooked. These factors necessitate continued attention for many years to come. It is important to note that the idea and methodology employed in this study extend beyond COVID-19 and encompass other events, especially those occurring at the edge of time series, which can present challenges for standard breakpoint analyses.

These examples and scenarios should be considered within the context of rigorous plausibility assessment and underlying assumptions. To ensure the validity of the scenario-based results, collaboration with epidemiological experts is recommended. This is crucial for the reliability and robustness of the analysis in dealing with the complexity associated with such events. The confrontation of our setting with the expert opinion in literature, e.g. by \cite{telenti2021after}, shows an agreement to a significant extent.

In the research paper, the impact of COVID-19 on mortality rates is incorporated into the mortality model using a binary indicator variable called $covid_i$. This variable is specific to each country $country_i$ and takes the value of 1 for the years 2020 and 2021, representing the period during which the COVID-19 pandemic had a significant impact. For all other years outside this period, $covid_i$ takes the value of 0. The values taken for future predictions are subject to the scenario assumptions elaborated below.

The introduction of this binary indicator variable allows the model to distinguish between the years affected by COVID-19 and the years that are not. By including $covid_i$ in linear interaction with the country-specific effects, the model can capture the country-specific impact of the pandemic on mortality rates during the years 2020 and 2021. This means that the model is able to estimate how the mortality rates deviated from the expected trend in these two years due to the influence of COVID-19.:

\begin{equation}
\begin{aligned}\label{eq:first}
\log(\mu_{apc,i}) = \beta_0 + f_{ap,s}(age_i, period_i) + \beta_{covid} covid_{i} * country_{i} + \log(E_{i})
\end{aligned}
\end{equation}

\vspace{0.2cm}
\[ covid_{i} = 
    \begin{cases} 
    \mbox{$1$,} &  \mbox{for $ period_i \in \{2020, 2021\} $} \\ 
    \mbox{$0$,} &  \mbox{for $ period_i \leq 2019$}
    \end{cases} \]
\vspace{0.2cm}

Four different scenarios are defined to model the COVID-19 impact on the mortality rates:

\hspace*{-0.6cm}\textbf{Scenario I: }In this scenario, the assumption is made that COVID-19 will disappear in the future. The model is trained using data up to 2019 only, excluding the years 2020 and 2021. The predictions are then made for the years 2020-2025, assuming no long-term effects of COVID-19 on mortality. This approach treats COVID-19 as a special event that does not have any influence on mortality in the upcoming years. The model focuses on the underlying mortality trend without considering the impact of COVID-19 and thus without the covid-indicator.

\hspace*{-0.6cm}\textbf{Scenario II: }In Scenario II, the expectation is that the full COVID-effect will persist in the future. The model is trained using mortality data up to and including 2020, which includes the years significantly affected by COVID-19. Predictions are made for the next years, assuming that the COVID-19 indicator variable $covid_{i}$ remains set to 1, indicating the presence of COVID-19. This scenario assumes that the COVID-related situation will continue similarly as it did in 2020 and 2021, and that it will have a consistent effect on mortality over the coming years.

\hspace*{-0.6cm}\textbf{Scenario III: }In this scenario, the assumption is made that the COVID-effect will flatten over time. Similar to Scenario II, the model is trained using mortality data up to and including 2021. However, in this case, the COVID-19 effect is assumed to decrease exponentially over time. The predictions take into account the diminishing impact of COVID-19 in the future, reflecting the belief that the effect of COVID-19 on health and mortality will slowly flatten out and eventually disappear after a few years. Therefore, the $covid_{i}$ indicator takes exponentially decreasing values between 1 and 0 for each year.  

\hspace*{-0.6cm}\textbf{Scenario IV: }In Scenario IV, the focus is on adjusting for excess mortality associated with COVID-19. The years 2020 and 2021 are treated as outliers, but the excess mortality is explicitly considered. The model calculates the difference between the expected death counts and the actual mortality counts for these two years to account for the excess mortality. It is assumed that the excess mortality will not average out over the coming years and must be explicitly accounted for. The baseline mortality, representing the mortality trend without the influence of COVID-19, remains unchanged. This scenario allows for separate consideration of the excess mortality caused by COVID-19 while keeping the baseline mortality unchanged.

These different scenarios and their technical implementations provide insights into the potential future trajectories of mortality, taking into account various assumptions about the impact of COVID-19.  

\begin{figure}[hbt!]
\centering
\includegraphics[width=0.75\linewidth]{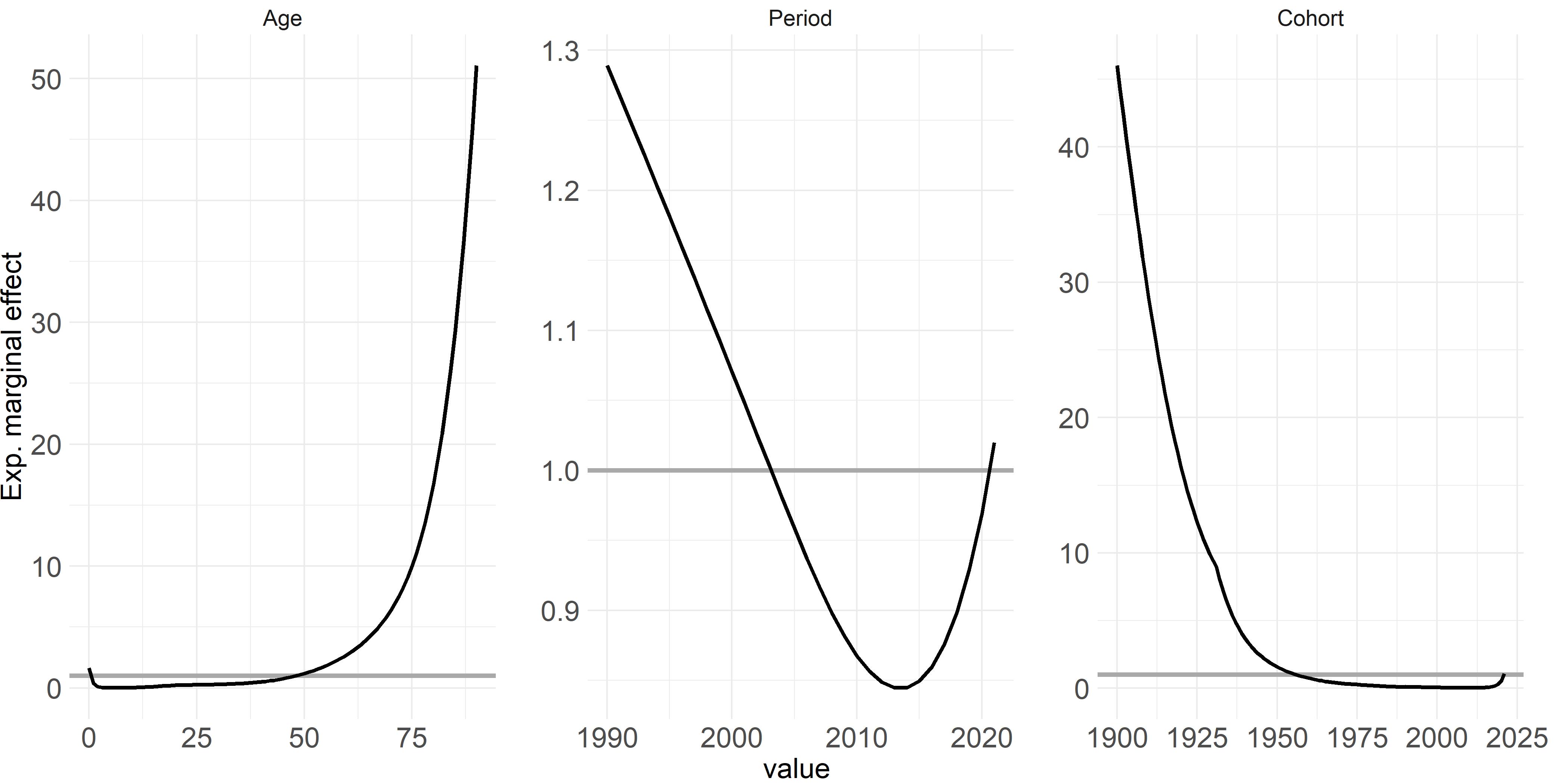}
\caption[Estimated marginal effects APC (1990-2020)]{The figure illustrates the estimated marginal effects of age, period, and cohort on mortality rates across multiple countries and genders. The horizontal lines in the plot indicate the level of no effect. The GAM model used in this analysis was fitted using data from the years 1990 to 2015 and covers ages ranging from 0 to 90.}
\label{fig_4}
\end{figure}

Figure \ref{fig_4} depicts the effects of the model based on age, period, and cohort effects. The period effect shows a significant increase up to the year 2020, reflecting the heightened impact of that specific year on mortality rates. The analysis reveals that the age and cohort effects align with expectations from the previous model until 2015. However, the period effect exhibits a sharp rise, particularly leading up to 2020, indicating a significant impact of the year on mortality rates. Specifically, the progress made in improving mortality rates over the past years or even decades appears to have been counteracted by the effect of COVID-19, resulting in a regression to the level seen around 2003.

\begin{figure}[hbt!]
\centering
\includegraphics[width=0.9\linewidth]{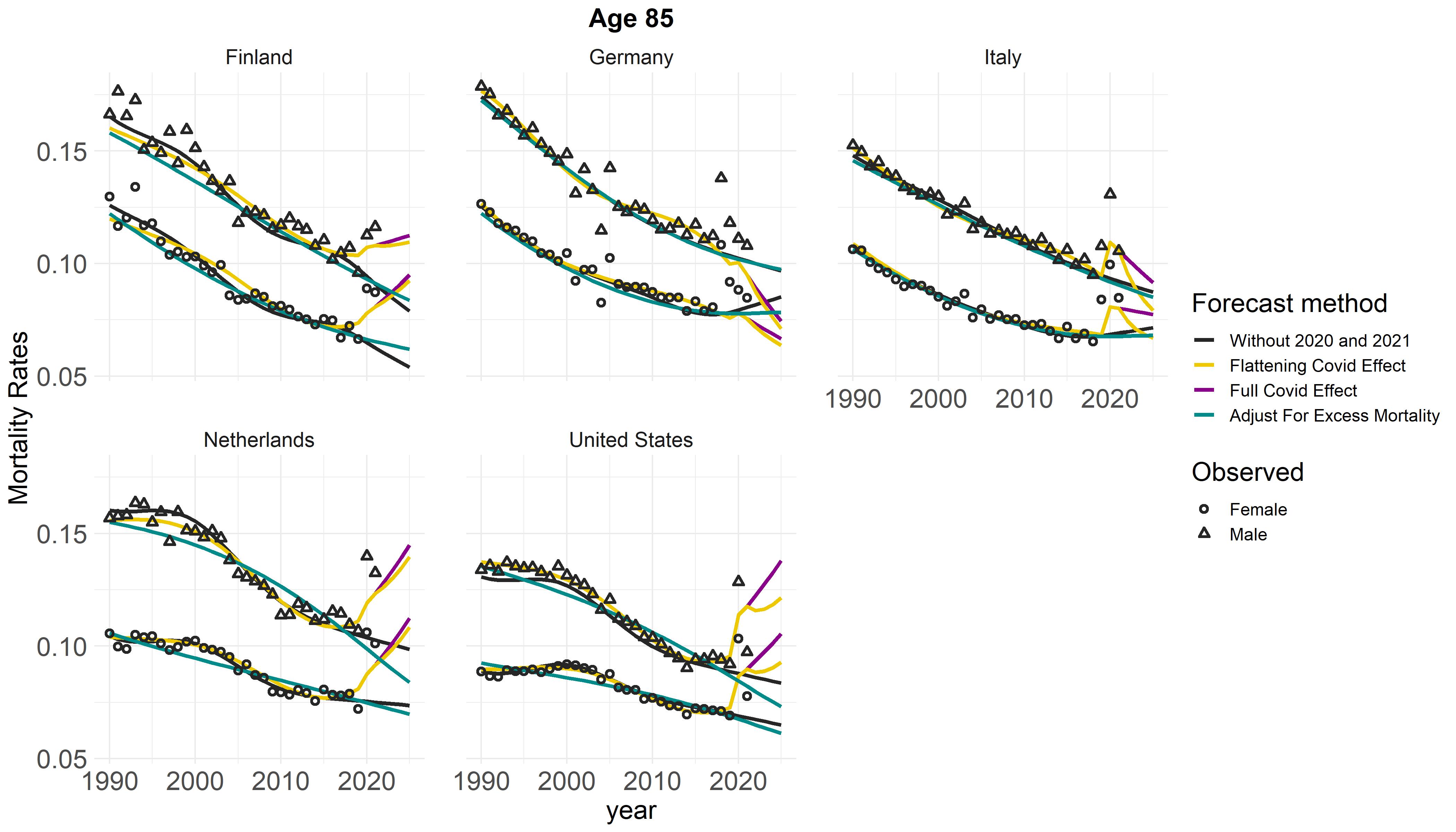}
\caption[Trend forecasts according to the scenarios for 85-year-olds]{The figure displays various trend forecasts for 85-year-olds based on different scenarios. The adjusted time period spans from 1990 to 2021, with an additional adjustment for excess mortality in the scenario requiring it, extending the period to 1990-2022. The forecast extends beyond the observed data to the year 2025.}
\label{fig_5}
\end{figure}

Figure \ref{fig_5} illustrates the outcomes of the four scenarios for different countries and genders, focusing on 85-year-olds. The observed mortality rates are compared with the estimated rates. Outliers can be observed for Italy, the US, and the Netherlands, particularly for the years 2020 and 2021, indicating the more pronounced impact of COVID-19 in these countries. Different trend forecasts in the scenarios capture the varying effects of COVID-19 on mortality rates, with Scenario IV considering excess mortality. The image highlights the noticeable difference between the two extreme scenarios, 1 and 2, particularly evident in the case of Italy. Scenario 1 represents a milder assumption, while Scenario 2 depicts a more severe projection. It is important to emphasize that the future forecasts vary for different countries and age groups, influenced by their respective past behaviors, as well as political and medical responses to the pandemic.

In summary, forecasting future mortality rates presents challenges due to the unavailability of exact quantitative evaluation. However, by discussing the trends and assumptions of the forecast in relation to the scenarios, it becomes possible to assess the plausibility of the predictions. The four alternative scenarios considered in this analysis represent a range of expected impacts, spanning from mild to severe. To ensure a rigorous assessment, these scenarios and their underlying assumptions were thoroughly evaluated and discussed in collaboration with epidemiological experts. This approach including the content of scenarios aligns with existing literature and enhances the credibility of the forecast analysis \autocite{telenti2021after}. The strength of this approach lies in its comprehensive evaluation of different mild to severe scenarios, which inherently constitutes an uncertainty analysis of the results presented. By testing various scenarios that span a range of possible mortality outcomes, the study acknowledges and addresses the uncertainties inherent in mortality forecasting, particularly in the context of the COVID-19 pandemic. This approach allows for a holistic understanding of the potential trajectories of future mortality trends under different conditions, providing valuable insights to policymakers and stakeholders in planning and decision-making. The robust consideration of multiple scenarios reflects our commitment to ensuring a thorough assessment of the model's performance and its adaptability to diverse and uncertain circumstances, enhancing the credibility and applicability of the proposed methodology.

\section{Conclusion}

To summarize, this research work focused on addressing the challenge of capturing the COVID-19 effect in future mortality forecasting. Two methods were discussed to improve modeling and forecasting.

Firstly, the study compared Generalized Additive Models (GAM) and Tree-Based Machine Learning (ML) techniques with traditional approaches. ML techniques revealed weaknesses in traditional models and bridged the gap between standard and modern explanations. GAM provided interpretable marginal effects, and both techniques significantly enhanced fitting performance. GAM however demonstrated superior forecast accuracy.

Secondly, in terms of practical application, the study applies the GAM in the APC framework with penalized smoothing second-order splines to forecast future mortality trends in a cross-country fashion. It considers the potential extension of the model to incorporate socio-economic variables, which is a novel approach. To compare the performance of different approaches and assess the effects of COVID-19 on mortality rates, the research conducts a detailed examination of five countries: Germany, Finland, the Netherlands, Italy, and the United States. It utilizes data from the Human Mortality Database, supplemented with COVID-related data collected on a weekly basis and aggregated to a yearly level. The study defines four future scenarios to facilitate trend forecasting and provide insights into the potential impact of COVID-19 on mortality rates.

Overall, this work contributes to the existing literature by introducing enhanced models, comparing different approaches, and providing insights into the future development of mortality rates, considering the impact of COVID-19, in a cross-country context. The specific contribution of the GAM approach with the APC framework in this research lies in its novel application for mortality trend forecasting, particularly in the context of incorporating the impact of COVID-19, to be broken down into the following contribution pieces: \\

\hspace*{-0.6cm}\textbf{Flexibility and Nonlinear Modeling}: GAMs are a well-established model class that extends Generalized Linear Models by incorporating non-parametric smooth components. This property makes them well-suited for modeling nonlinear effect structures, which is essential when dealing with mortality data that often exhibits complex relationships between age, period, and cohort. 

\hspace*{-0.6cm}\textbf{Incorporating COVID-19 Effect}: By applying the GAM within the APC framework, the model can effectively capture the impact of COVID-19 on mortality trends. The inclusion of COVID-19 as an additive factor allows the model to account for the unprecedented disruption caused by the pandemic on mortality rates, leading to more accurate and meaningful forecasts.  

\hspace*{-0.6cm}\textbf{Cross-Country Forecasting}: The GAM with APC framework enables cross-country forecasting of future mortality trends. This means that the model can be applied to multiple countries, providing insights into how different nations are affected by both general trends in mortality improvement and the specific impact of COVID-19. This capability is crucial for understanding variations in mortality patterns across different regions.  

\hspace*{-0.6cm}\textbf{Possible Inclusion of Socioeconomic Variables}: Another significant contribution of the GAM with APC framework is its potential for further extensions. It can accommodate the inclusion of socioeconomic variables as additional factors, allowing for a more comprehensive understanding of mortality trends. This feature is particularly important as socioeconomic factors can significantly influence mortality rates and can be useful for policy and decision-making purposes. 

\hspace*{-0.6cm}\textbf{First-Time Combination}: To our knowledge, this research is the first to combine the GAM approach within the APC framework, which focuses on incorporating the effects of COVID-19 into mortality modeling. This novel combination enhances the state-of-the-art by leveraging the strengths of both GAM and APC to tackle the unique challenges posed by the pandemic's impact on mortality rates. \\

Despite the waning impact of COVID-19 at present, it is crucial to recognize the enduring importance of historical data and the persisting uncertainties that lie ahead. These factors emphasize the need for ongoing attention in the years to come. It is important to acknowledge that the concept and methodology utilized in this study extend beyond COVID-19, encompassing other events that occur at the fringes of time series data. Such events can pose challenges for conventional breakpoint analysis methods. Looking towards future research directions, the availability of full data for several COVID-19 impacted years in the updated STMF database presents an exciting opportunity for further analysis. Researchers can now conduct comprehensive tests to determine the most likely scenario and assess the accuracy of the model's parameters in light of this new data, providing valuable insights into mortality forecasting amidst the COVID-19 pandemic. Moreover, this research highlights the broader significance of incorporating expert knowledge when considering extreme event impact assessment in mortality analysis beyond the current pandemic context. These findings have implications not only for the longevity insurance industry but also underscore the need for scenario analysis in mortality forecasting to better prepare for and understand the potential impacts of various extreme events on population health.


\printbibliography

\appendix

\section{Age-Period-Cohort Model}

The Age-Period-Cohort (APC) model extends the LC model by including a cohort effect $\gamma_{t-a,s}$ and omitting the age-specific improvement rates. The cohort is generally computed by $cohort = year - age$. The model predictor has the following expression: 

\vspace{0.2cm}
\begin{equation}
\label{eqn:apc}
\begin{aligned}
\eta_{a,t,s} = \beta_{a,s} + \kappa_{t,s} + \gamma_{t-a,s}
\end{aligned}
\end{equation}
\vspace{0.2cm}

The application of this model has its origins in the field of medicine and demography and goes back a long way (\autocite{clayton1987models}; \autocite{hobcraft1985age}). However, \cite{currie2004smoothing} was the first who considered this type of model in the actuarial field. With the Poisson distribution assumption and the log link function remaining the same, it can be traced back to the general shape of Generalized Age-Period-Cohort models \autocite{villegas2015mortality}. The identifiability can be ensured with the following constraints: $ \sum_{t} \kappa_{t,s} = 0$, $\sum_{c = t_{min} - 90}^{t_{max} - 0} \gamma_{c,s} = 0$, $\sum_{c = t_{min} - 90}^{t_{max} - 0} c \gamma_{c,s} = 0$, indicating that the cohort effect oscillates around zero, with no apparent linear trend. \\ %

\section{Lee Carter}

Lee Carter is a model for estimating mortality rates with the following assumptions: 

\vspace{0.2cm}
\begin{equation}
\begin{aligned}
D_{a,t,s} \sim Poisson(E_{a,t,s} \cdot \mu_{a,t,s}), \: \text{independent distributed} 
\end{aligned}
\end{equation}

\begin{equation}
\begin{aligned}
\eta_{a,t,s} = \alpha_{a,s} + \beta_{a,s}\kappa_{t,s}, \\
\text{with} \;\; \log \mu_{a,t,s} = \eta_{a,t,s}
\end{aligned}
\end{equation}
\vspace{0.2cm}

Thus, fitting a Lee Carter model means basically to provide estimates for mortality rates using the two inputs exposure $E_{a,t,s}$ and death counts $D_{a,t,s}$. These estimates will be marked as $\mu_{a,t,s}^{LC}$ in this subsection. The expected number of deaths according to the Lee Carter fit can be calculated as follows:  

\vspace{0.2cm}
\begin{equation}
\begin{aligned}
D_{a,t,s}^{LC} = E_{a,t,s} \cdot \mu_{a,t,s}^{LC}
\end{aligned}
\end{equation}
\vspace{0.2cm}

\section{Decision Trees}

Decision trees are non-parametric supervised learning methods that can be applied to both regression and classification problems, aiming for a model to predict the outcome variable by learning simple decision rules. These rules are implied from the features in the training data. They are based on the recursive rectangular partitions of the feature space via a sequence of binary splits and can be seen as a piecewise constant approximation of the underlying true function.  During training, observations are forwarded along the resulting tree structure until they end up in a particular leaf node. This set of splitting rules is summarized in a tree \autocite{hastie2009elements}. Predictions of the response for a given observation with a certain feature space are made by using the average of the training observations in the region that the observation falls into \autocite{james2017islr}.

Let $\Theta$ be the number of partitions and $R_1, R_2, \dots ,R_\Theta$ the terminal regions of the feature space. Then, the decision tree estimator yields in the following expression: 

\vspace{0.2cm}
\begin{equation}
\begin{aligned}
\hat{q}_{a,t,s} =  \sum_{\theta = 1}^{\Theta} \bar{q}_{a,t,s} \mathds{1}_{\{ (a,t,s) \in R_\theta \}}   
\end{aligned}
\end{equation}
\vspace{0.2cm}

where $\bar{q}_{a,t,s}$  is the prediction for a given observation in leaf node $\theta \in \Theta$. The algorithm is designed in a way, that each split is loss-minimal across all child nodes, requiring full evaluation of all possible split and threshold configurations. For further details on the Classification and Regression Trees (CART) algorithm, the reader is referred to \cite{gordon1984classification}. To implement decision trees in R the package \texttt{rpart} was used \autocite{therneau2015package}. 

In general, trees are easy to interpret and have a facile handling of all feature types, including interactions (the latter ability also holds for Random Forest). They are able to fit perfectly every pattern in the training data if allowed for full growth. Overfitting issues are posed, since they can lack robustness and are highly sensitive to data modifications. Thus, probably the most important drawback of Decision Trees is its low-bias and high-variance nature. 

\section{Gradient Boosting}

Gradient Boosting is another form of an ensemble learner that is based on the weighted combination of weak predictive learners such as Decision Trees, usually outperforming Random Forest \autocite{hastie2009elements}. The model is built stepwise and optimized by a differentiable loss function, minimizing the in-sample loss \autocite{hastie2009elements}. It builds the model stepwise, like other boosting methods, and generalizes them by allowing optimization of any differentiable loss function. Whereas in bagging multiple samples of the original training dataset are used to fit a separate decision tree to each one independent of the others and to combine all trees into a single predictive model, boosting grows the trees sequentially, meaning the information gained from the previous trees is used to grow the current one. 
This helps to overcome the major issue of training a single large Decision Tree by possibly resulting in an overfitting problem. The gradient boosting algorithm instead learns by constructing a new model based on the previous one and adding the $i^{th}$ base learner $h^{(i)}_{a,t,s}$:

\vspace{0.2cm}
\begin{equation}
\begin{aligned}
\hat{q}^{(i)}_{a,t,s} = \hat{q}^{(i-1)}_{a,t,s} + \lambda_{i}h^{(i)}_{a,t,s}
\end{aligned}
\end{equation}
\vspace{0.2cm}

The model will be improved in such a way that the current residual will be used as an outcome to fit a new Decision Tree and to add this into the originally fitted function with the notion to update the residuals. So, the gradient boosting algorithm fits the new predictor to the residual errors made by the previous predictor. The shrinkage parameter $\lambda_i$ helps to run the process even slower allowing for more trees and more detailed enhancement of the residuals. Overall in contrary to the bagging methodology, each tree depends on the previous ones \autocite{deprez2017machine}.
Even though the gradient boosting keeps on minimizing the errors, this can cause overfitting in case of a lot of noise in the data and is computationally time and memory expensive, especially because trees are built sequentially (not in parallel as the Random Forest do). Due to the high flexibility, the gradient boosting algorithm also tends to be harder to tune than Random Forest \autocite{hastie2009elements}.
The package \texttt{gbm} was utilized to implement this algorithm in R \autocite{ridgeway2004gbm}. The key parameter configurations were set up following the recommendations of \cite{deprez2017machine} and \cite{levantesi2019application} and will be described in detail in the analysis section.
\end{document}